\documentclass[journal]{IEEEtran}
\usepackage{amsmath,graphicx}
\usepackage{amssymb}  
\usepackage{amsthm}

\newcommand{\tb}{\textbf}

\title{Complex tensor factorisation with PARAFAC2 for the estimation of brain connectivity from the EEG}
\author{Loukianos Spyrou, Mario Parra, and Javier Escudero\thanks{This work was supported by EPSRC, UK, Grant No. EP/N014421/1.}}

\begin{document}
%
\maketitle
\begin{abstract}
Objective: The coupling between neuronal populations and its magnitude have been shown to be informative for various clinical applications. One method to estimate brain connectivity is with electroencephalography (EEG) from which the cross-spectrum between different sensor locations is derived. We wish to test the efficacy of tensor factorisation in the estimation of brain connectivity. 
Methods: Complex tensor factorisation based on PARAFAC2 is used to decompose the EEG into scalp components described by the spatial, spectral, and complex trial profiles. An EEG model  in the complex domain was derived that shows the suitability of PARAFAC2. A connectivity metric was also derived on the complex trial profiles of the extracted components. 
Results: Results on a benchmark EEG dataset confirmed that PARAFAC2 can estimate connectivity better than traditional tensor analysis such as PARAFAC within a range of signal-to-noise ratios. The analysis of EEG from patients with mild cognitive impairment or Alzheimer's disease showed that PARAFAC2 identifies loss of brain connectivity better than traditional approaches and agreeing with prior pathological knowledge. \\
Conclusion: The complex PARAFAC2 algorithm is suitable for EEG connectivity estimation since it allows to extract meaningful coupled sources and provides better estimates than complex PARAFAC.\\
Significance: A new paradigm that employs complex tensor factorisation has demonstrated to be successful in identifying brain connectivity and the location of couples sources for both a benchmark and a real-world EEG dataset. This can enable future applications and has the potential to solve some the issues that deteriorate the performance of traditional connectivity metrics.
\end{abstract}
\section{Introduction}
\label{sec:intro}

Tensor factorisation has found many applications in several areas such as antenna array processing, blind source separation, biomedical signal processing, feature extraction, and classification \cite{Bro1997,Harshman1970,Harshman1972,comon2009tensor}. A tensor is a multi-way representation of data or a multidimensional array. Each dimension in the tensor is called a mode or a way. Using tensor factorisation, the true underlying structure of that data can be revealed. Tensor factorisation methods have been shown to be powerful for describing signals which in general change in time, frequency, and space. Tensor analysis can provide a good way to discover the main features of the data and extract the hidden underlying information especially in the case of having big data size. 

Several tensor based methods have been suggested for decomposition and multi-way representation of data. The PARAFAC decomposition \cite{Bro1997,Harshman1970,kroonenberg1980principal} is one of the common tensor factorisation methods which is a generalisation of singular value decomposition (SVD) to higher order tensors. Using the PARAFAC model, the data are decomposed into a sum of rank-1 tensors of lower dimensions than the original data. Therefore, as suggested in \cite{cichocki2013tensor}, it can be employed to compress the high dimensional data and extract their significant features. 

The application of tensor decomposition can be significant for biomedical signals, such as EEG \cite{SaneiB}, where many transient events and movement related sources and artifacts are involved and most sources are inherently nonstationary. Moreover, the related brain neural process exhibit specific space-time-frequency locations.  EEG signals in particular, consist of multichannel recordings with good temporal resolution which subsequently offers good time-frequency resolution. The application of tensor analysis is then logical and the data can be factorised into its space, time and frequency modes \cite{Acar2007,Lee2007a,Spyrou2015b,Cong2015}. Tensor factorisation has been also applied to multi-subject data where the data can be factorised in the group level, identifying the common components \cite{Spyrou2015b,Spyrou2016jsps}. 

The coupling between neural processes has been investigated for various mental tasks such as attention, spatial navigation, perceptual binding and memory \cite{Nolte2004,Haufe2010,Kaminski1991,Haufe2013}. The activity in different areas can be phase-coupled, i.e., display systematic phase-delays over trials, a phenomenon called phase-synchronization, which has been hypothesized to be an important mechanism for creating a flexible communication structure between brain regions. However, it is well-known that indexing phase-synchronization through scalp-sensor (scalp-electrode) measurements can be complicated by five problems (see also \cite{VandeSteen2016}): (i) the choice of reference electrode \cite{Stam2007,Chella2016}, (ii) volume-conduction of source activity \cite{Brunner2016}, (iii) the presence of noise sources \cite{Vinck2011}, (iv) sample-size bias \cite{Vinck2011}, and (v) the coupling between sensors may not be due to the activity closest to those electrodes \cite{Sirota2008,Vinck2011}. These problems can spuriously inflate phase-synchronization indices and are partially overcome with the currently state-of-the-art methods employing the imaginary part of coherence as a basic step in their algorithms. These algorithms optimise phase-synchronisation measurements by minimising the effect of volume conduction and the presence of noise sources. However, they suffer from the limitation of reduced performance in the presence of noise sources and sample-size bias.

Tensor factorisation has been employed in brain connectivity studies primarily with the aim of dimensionality reduction or detection of dynamic changes \cite{Mahyari2014,Britta2015,Abasolo2008,Leonardi2013}. In this work, we extend the traditional tensor factorisation framework by considering complex valued tensors and we derive a connectivity metric based on the resulting factors. Complex tensors have been also used in factorisation schemes in \cite{Morup2008,VanderMeij2015,Kouchaki2015b}. In \cite{Morup2008} a multilinear decomposition is performed that explicitly models phase shifts between trials while in \cite{VanderMeij2015} phase shifts are also considered between electrodes, a phenomenon which arises in electrocorticography data. Unlike those studies we do not only consider phase shifts but a general EEG model which requires complex trial activations. Subsequently, the proposed methodology enables the estimation of connectivity between the components of the decomposition. This is accomplished by the decomposition of the EEG into channel, frequency and complex trial components. To this end, we derive an EEG model in the complex domain and show that PARAFAC2 is better suitable than PARAFAC in factorising that model and provide a metric that estimates the coupling between sources. In section \ref{sec:EEG}, we describe the EEG model in the complex domain and together with an introduction on connectivity measures. Section \ref{sec:tensor} formulates the factorisation procedure and we demonstrate the suitability of PARAFAC-2 and a connectivity metric that is defined on the extracted components. In section \ref{sec:results} we show results on a benchmark connectivity dataset and real Alzheimer's EEG data. Section \ref{sec:discussion} puts the benchmark results into context, showcasing the conformity of the real AD data to prior pathological knowledge and the Scaffolding Theory of Cognitive Aging \cite{Cabeza2002}. Section \ref{sec:conclusions} concludes the paper.

\section{EEG model}\label{sec:EEG}

The EEG measures the concurrent activity from multiple neural sources mixed into multiple sensors. Various forward models have been developed that map the way that such sources are propagate and are mixed into the scalp electrode sensors \cite{Hallez2007,Mosher1999}. Each source usually describes a separate mental process or group of related processes \cite{SaneiB} and estimating their properties has been attempted in a variety of ways \cite{Michel2004,Spyrou2008,Cong2015}. 

EEG is traditionally analysed with respect to channels and temporal samples resulting in a model:

\begin{equation}
\tb{X}(t)=\tb{A}\tb{S}(t)=\sum\limits_{i=1}^n \tb{a}_i  s_i(t)
\end{equation}
where $\tb{A} \in \mathbb{R}^{m\times n}$ describes the forward model of $n$ sources on $m$ electrodes and $\tb{S}(t)\in \mathbb{R}^{n}$ the source matrix where each source has a duration of $T$ temporal samples. EEG sources are commonly modelled as $k_{th}$ order autoregressive (AR) processes \cite{Pardey1996}:

\begin{equation}
	s_i(t)=\sum\limits_{\tau=1}^k h_i(\tau) w_i(t-\tau)
\end{equation}
where $w_i(t)$ is white gaussian noise and $h_i(t)$ the AR process parameters for the $i_{th}$ source. The parameters $h(t)$ can generate a variety of source types such as narrowband and lagged sources. When two sources are connected this can be modelled by bivariate AR processes such as \cite{Haufebb}:  
\begin{equation}\label{eq:bAR}
\left[ \begin{array}{c} s_i(t) \\ s_j(t) \end{array} \right] =\sum\limits_{\tau=1}^T \begin{bmatrix} h_{i}(\tau) & h_{ij}(\tau) \\ h_{ji}(\tau) & h_{j}(\tau) \end{bmatrix} \left[ \begin{array}{c} w_i(t-\tau) \\ w_j(t-\tau) \end{array} \right]
\end{equation}
The coefficients $h_{ij}$ describe the connection between the $i_{th}$ and $j_{th}$ sources also allowing for directionality in the connectivity when $h_{ij}\neq h_{ji}$.

Alternatively, each source may be considered in the frequency domain since many neural sources are oscillatory in nature \cite{David2006}. The mixing model remains the same since frequency transforms are linear functions of the time domain signal. Note that the transformed sources $s_i(f)$ are now complex valued describing the power and phase as a function of frequency.

In the frequency domain the EEG model is written as:
\begin{equation}
\tb{X}(f)=\tb{A}\tb{S}(f)=\sum\limits_{i=1}^m \tb{a}_i  s_i(f)
\end{equation}
where $\tb{X}(f) \in \mathbb{C}^{m\times F}$ and $s_i(f) \in \mathbb{C}^{1\times F}$ with $F$ the total number of frequency bins. Coupled sources are written as:
\begin{equation}
	\left[ \begin{array}{c} s_i(f) \\ s_j(f) \end{array} \right] = \begin{bmatrix} h_{i}(f) & h_{ij}(f) \\ h_{ji}(f) & h_{j}(f) \end{bmatrix} \left[ \begin{array}{c} w_i(f) \\ w_j(f) \end{array} \right]
\end{equation}

Due to the low signal-to-noise ratio of EEG signals, datasets usually consist of multiple measurements of the same state or task that generates the sources of interest. These measurements are called trials and aid in increasing the SNR since undesired source activities are considered uncorrelated to the activities of interest. Similarly, in a continuous recording, data are segmented into non overlapping windows, again called trials, in order to facilitate analysis in shorter time segments. The model is then written as:

\begin{equation}
\tb{X}(f,k)=\tb{A}\tb{S}(f,k)=\sum\limits_{i=1}^m \tb{a}_i  s_i(f,k)
\end{equation}
where $s_i(f,k)$ is the complex source activation at the $f_{th}$ frequency bin and $k_{th}$ trial. With the source model written as:

\begin{equation}
	s_i(f,k)=h_i(f) w_i(f,k) = p_i(f)y_i(f,k)
\end{equation}

Such a model allows for time-shifts between trials. Also, it allows to separate the frequency profile, $p_i(f)$, and trial to trial variations, $y_i(f,k)$, of each source resulting in:
\begin{equation}\label{eq:ceeg}
\tb{X}(f,k)=\sum\limits_{i=1}^m \tb{a}_i p_i(f) y_i(f,k)
\end{equation}
where $p_i(f) \in \mathbb{C}^F$ describes the complex amplitude of the $i_{th}$ source at frequency $f$. The term $y_i(f,k)$ describes the trial to trial variations in amplitude and phase. Such a model allows for typical narrowband EEG sources such as alpha sources with sparse $p_i(f)$. It is important to notice that $y_i(f,k)$ is both a function of frequency and trial.

For sources that are not connected $p_i(f)=h_i(f)$ and $y_i(f,k)=w_i(f,k)$ which are the frequency transforms of the AR model and white noise realisation respectively. For a connected source such as:
\begin{equation}
s_j(f,k)=h_j(f)w_j(f,k)+h_{ij}(f)w_i(f,k)
\end{equation}
the $p_i(f)$ and $y_i(f,k)$ describe more complicated relationships between the spectra and activations of the two sources. 

\subsection{Connectivity estimate}

Consider two electrodes $x_i$ and $x_j$, the coupling between them can be estimated by a variety of methods which use the cross-spectrum as an initial step:

\begin{equation}
Y_{i,j}(f,k)=x_i(f,k) x_j(f,k)^*
\end{equation}
which is the cross-spectrum between the electrodes at frequency $f$ and trial $k$. Subsequently, coupling is assumed to exist if an electrode is leading (or lagging) in phase consistently over trials. Several measures have been developed one of which is the phase lag index (PLI):

\begin{equation}
\Psi_{ij}(f) =|\mathbb{E} \{sign(\Im(x_i(f) x_j(f)^*))\}|
\end{equation}
where the expectation is taken over trials. If the electrode signals are of consistently different phase then $\Im(x_1(f) x_2(f)^*)$ will be of the same sign and $\Psi(f)$ will tend to 1. Measuring the brain connectivity using the PLI is optimal only for a pair of coupled sources as shown in \cite{Vinck2011}. For more than 2 coupled sources, cross-source interference arises and the optimality of the measure is lost in terms of volume conduction accuracy of spatial location. The PLI was chosen as the metric of interest since it is widely used in neuroimaging studies \cite{Fraschini2016,Stam2007,Kasakawa2016}.

\section{Complex tensor factorisation} \label{sec:tensor}

\subsection{Tensor model}

In order to alleviate the aforementioned issues with simple connectivity measures we propose a tensor factorisation procedure where the EEG is decomposed into distinct components in the complex domain. We consider a tensor EEG model, directly resulting from Equation \ref{eq:ceeg}, as a sum of components where each one is described by a triplet of real activations over electrodes $\tb{a}$, over complex frequencies $\tb{p}$, and complex trials $\tb{y}$ as:

\begin{equation}\label{eq:parafac2}
\begin{aligned}
& \tb{{X}}(f) = \tb{A}\tb{P}(f)\tb{Y}(f)\\
&= \tb{A} \begin{bmatrix} p_1(f) & \\  & \ddots \\ &&p_m(f)\end{bmatrix}\begin{bmatrix} \tb{y}_{1}(f) &\ldots & \tb{y}_m(f)  \end{bmatrix}
\end{aligned}
\end{equation}
where $\tb{{X}}(f) \in \mathbb{C}^{m\times F \times K}$. $\tb{y}_i(f) \in \mathbb{C}^{K\times 1}$ are the complex activations of the $i_{th}$ source over $K$ trials placed in the matrix $\tb{Y}$. $\tb{P} \in \mathbb{C}^{n \times n}$ is a diagonal matrix holding the complex frequency profiles of the $m$ sources in the diagonal for a single frequency $f$. This model is the PARAFAC2 model. If the source activations were identical for all frequencies i.e. $s_i(f_n)=s_i(f_m)$ then the model would correspond to the typical PARAFAC model with:
\begin{equation}
\begin{aligned}
&\tb{{X}}(f)=\tb{A}\tb{P}(f)\tb{Y}\\
&=\tb{A} \begin{bmatrix} p_1(f) & \\  & \ddots \\ &&p_m(f)\end{bmatrix}\begin{bmatrix} \tb{y}_{1} &\ldots & \tb{y}_m  \end{bmatrix}
\end{aligned}
\end{equation}
There are two reasons that the source activations are dependent on the frequency: a) due to $w_i(f,k)$ being a complex source amplitude it achieves different complex values for different frequencies and trials, and b) differences between trials can be accommodated by considering that a variable process such as $w'_i(f,k)=d_i(f,k)w_i(f,k)$. For phase-shifts between trials we can consider: $d_i(f,k)=exp(-j\theta_kf)$ being also a function of frequency and trials. Hence, this formulation allows the decomposition of EEG into complex sources that are potentially not phase locked to a stimulus. 

\subsection{Connectivity estimation in component space}

It then becomes possible to measure connectivity information in component space by considering the trial activations. The connectivity measure is performed similarly to the cross-spectrum between electrodes but now is between components:

\begin{equation}\label{eq:tpli}
\Psi^s_{ij}(f) =|\mathbb{E} \{sign(\Im({ \tb{s}}_i(f) {\tb{s}}_j(f)^*)\}|
\end{equation}
which measures the coupling between component $i$ and $j$ where $\tb{s}_i(f)=p_i(f)\tb{y}_i(f)$. In order to obtain a connectivity profile over the scalp we can use that PLI value weight by the activation of the sources over the scalp:

\begin{equation}
\tb{C}^a_{ij} =\Psi^s_{ij} (\tb{a}_i\tb{a}_j^T+\tb{a}_j\tb{a}_i^T)
\end{equation}
which gives a symmetric connectivity profile of the coupling between the two sources. Furthermore, the frequency information can be incorporated by weighting by the mean frequency magnitude over a specific band of the desired sources:
\begin{equation}\label{eq:conn}
\tb{C}^{af}_{ij} =\Psi^s_{ij} (\tb{a}_i\tb{a}_j^T+\tb{a}_j\tb{a}_i^T)(\frac{1}{L}\sum\limits_{l}^L |\tb{s}_i(f_l)|+|\tb{s}_j(f_l)|)
\end{equation}

\subsection{Complex PARAFAC2}

The factorisation procedure entails alternating least squares minimisations of the following equation:

\begin{equation}\label{eq:parafac2alte}
\begin{aligned}
 & \underset{\tb{A},\tb{P}(f), \tb{Y}(f)}{\operatorname{argmin}} ||\tb{{X}}(f) - \tb{A}\tb{P}(f)\tb{Y}(f)||^2\\
\end{aligned}
\end{equation}
However, the solution to the factorisation of equation \ref{eq:parafac2alte} is not unique \cite{Kiers1999} and the PARAFAC2 model imposes an additional constraint to facilitate uniqueness. This is accomplished by having the cross-products of the $3_{rd}$ mode be constant over its index:
\begin{equation}
\tb{Y}(f_i)^H\tb{Y}(f_i)=\tb{K}, \,\,\forall i
\end{equation}
which is implemented by having $\tb{Y}(f)=\tb{Q}(f)\tb{H}$ where $\tb{Q}(f) \in \mathbb{C}^{K \times R}$ is an orthonormal matrix and $\tb{H} \in \mathbb{C}^{R \times R}$ any square matrix. That leads to the modified factorisation algorithm as:
\begin{equation}\label{eq:parafac2alte2}
\begin{aligned}
& \underset{\tb{A},\tb{P}(f), \tb{H}}{\operatorname{argmin}} ||\tb{Q}(f)^H\tb{{X}}(f) - \tb{A}\tb{P}(f)\tb{H}||^2\\
\end{aligned}
\end{equation}
which is solved by standard PARAFAC algorithms. Note that the minimisation over $\tb{A}$ is performed in the real domain.

The constraint is not optimal for EEG data since it enforces the complex inner product between sources to be equal for all frequencies which is not necessarily true. However, in our case where sources are coupled only in a narrow band (e.g. alpha sources), the inner product in the rest of the frequency range will be weighted by the activity on the 2nd mode. The results of the simulation also support the usefulness of the constraint in our setting (c.f. \ref{sec:sim})

\subsection{Experimental validation and application to real world signals}
 
\subsubsection{Validation with synthetic data}
The PARAFAC-2 algorithm is evaluated initially on the Berlin Brain Connectivity Benchmark \cite{Haufe}. This framework provides the capability of EEG data generation that resemble as much as possible real EEG data with customisable parameters. The simulation includes a realistic a 108-channel real head model and the generation of pseudo-EEG sources that can be either narrowband (e.g. alpha band) or pink noise sources corresponding to background brain activity. Furthermore, it enables the creation of coupled sources based on the bivariate AR model of Equation \ref{eq:bAR} or not coupled sources with zeros on the off-diagonals. The default settings provided in the framework correspond to a realistic EEG recording \cite{Haufe}. These include a pair of coupled sources with spread over the cortex resembling a brain network and a large number of background EEG activities. For the purposes of this study we desire to evaluate: a) the performance of the algorithm in estimating the coupled sources in terms of the power ratio (PR) of the coupled sources to the whole EEG power and b) the influence of the number of components of the tensor factorisation to the performance of the algorithm. We define three different metrics to accomplish this goal. Firstly, the explained variance (EV) of the tensor factorisation which describes how well the algorithm is fit to the data:
\begin{equation}\label{eq:erred}
EV=\frac{\sum\limits_f ||\tb{A}\tb{D}(f)\tb{Z}(f)||^2_F}{\sum\limits_f ||\tb{X}(f)||^2_F}
\end{equation}
Secondly, a connectivity estimation metric CONN as described in \cite{Haufe} which describes the ability of the algorithm to detect whether there exist coupled sources in an EEG segment or recording. A correct estimation is awarded with $+1$ while a wrong one with $-2$ with the chance level being at $-0.5$. Lastly, the LOC metric as described in \cite{Haufe} which estimates how well the algorithm can identify the coupled sources and their spatial profile in terms of the brain octant they belong to. The LOC metric awards for each source $+0.5$ when the octant is predicted correctly and $-0.5$ otherwise. Chance level of this metric is $-0.5$.

In summary the settings of the simulated EEG are:
\begin{itemize}
\item Channels: 108
\item Sampling frequency: 100Hz
\item Coupled sources: 2
\item Noise sources: 500
\item AR model order: 5
\item Power ratio (PR) range: 20$\%$ - 90 $\%$
\item Sensor noise: 10$\%$
\item Chance a dataset contains coupled sources: 50$\%$
\item Datasets for each PR: 100
\item Duration of each dataset: 3 minutes
\item Realistic head model lead-fields: 2000
\end{itemize}
Each dataset is sliced into 1 second trials resulting in a EEG of dimensions $108 \times 180$. A common average reference is applied and the data are also detrended and baselined. Furthermore, we tensorise the data with the short time fourier transform with $1Hz$ Hanning tapers in the range of $[1-40]Hz$. This results in an EEG of dimensions $108 \times 40 \times 180$ which is input to the PARAFAC2 and PARAFAC algorithms.  To avoid the effect of local minima we perform 10 runs of 10 initialisations each and retain the run with the highest coupling value. This is performed for both the datasets that contain coupled sources and the ones that do not. All pairs of the trial mode factors are then input into equation \ref{eq:conn} from which the pair with the highest coupling is selected to compute the CONN and LOC metrics. The CONN metric was estimated by finding the threshold that maximally separated the coupled datasets with the ones without coupling. The LOC was estimated by correlating the obtained spatial component of each of the paired sources with each of the lead-field of the head model and picking the octant with the highest correlation.

\subsubsection{Illustration in AD}

The algorithm was applied on two electroencephalography (EEG) datasets on a memory task from Alzheimer's patients and control subjects \cite{Pietto2016}. In dataset-1, there were 128-channel recordings from 13 patients with mild cognitive impairment (MCI) and 14 control subjects (mean age: 73) while for dataset-2 there were 64-channel recordings from 10 patients with MCI and carrying the gene of familial Alzheimer's disease (FAD) and 10 control subjects (mean age: 44). Similar preprocessing is performed in the EEG signals of the AD datasets resulting in tensors $128/64 \times 40 \times ~50$ with the number of trials slightly different between subjects with identical algorithm parameters as in the benchmark dataset.

\section{Results}\label{sec:results}

\subsection{Simulated EEG}\label{sec:sim}

\begin{figure*}[htbp]
	\centering    
	\includegraphics[width=18cm]{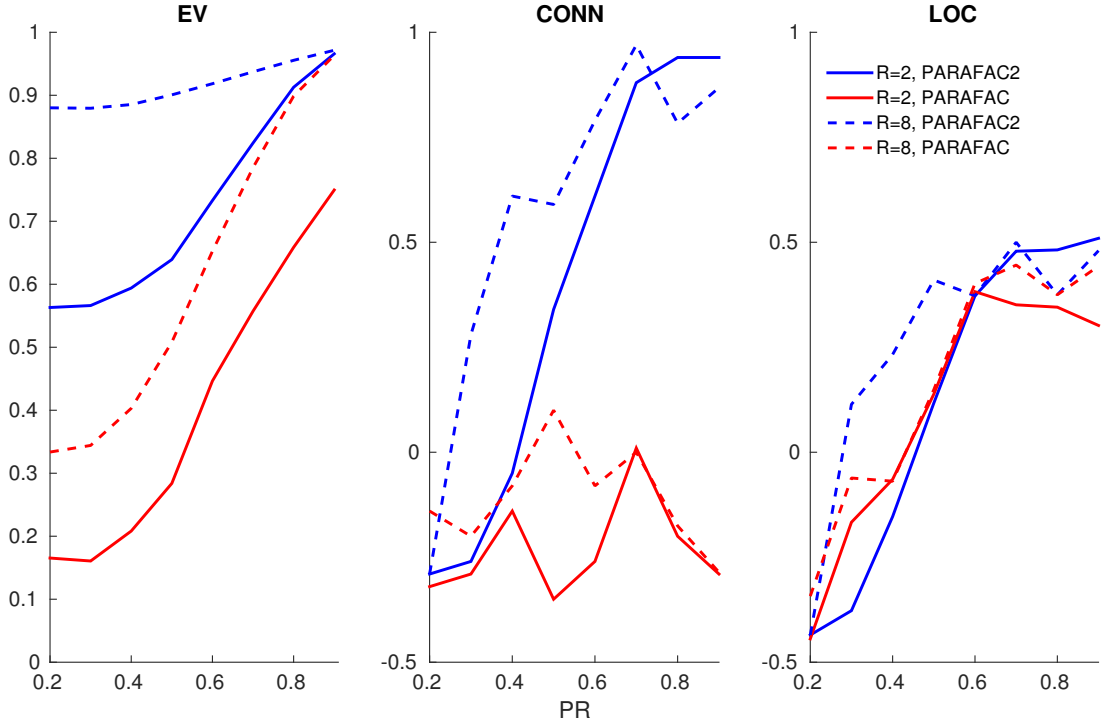}
	\caption{Explained variance (EV), CONN and LOC of the PARAFAC-2 algorithm for a range of PR values and different number of components. For each SNR value we generate 100 datasets where in each one the signal sources are in the alpha frequency band and the noise sources are pink noise realisations.}
	\label{fig:sim}
\end{figure*}

Firstly, we apply the complex tensor factorisation algorithm for a range of PR values and for $R=2$ and $R=8$. The number of components was set to a maximum of 8 to avoid non-uniquness issues. For each PR value we generate 100 datasets where in each one the signal sources are in the alpha frequency band and the noise sources are pink noise realisations. Then, we compute the three metrics and the results are seen in Figure \ref{fig:sim}. In \cite{Haufebb} a CONN value of 0.5 and a LOC value of 0.51 were obtained.

\subsection{Alzheimer's EEG}

\begin{figure*}[htbp]
	\centering    
	\includegraphics[width=18cm]{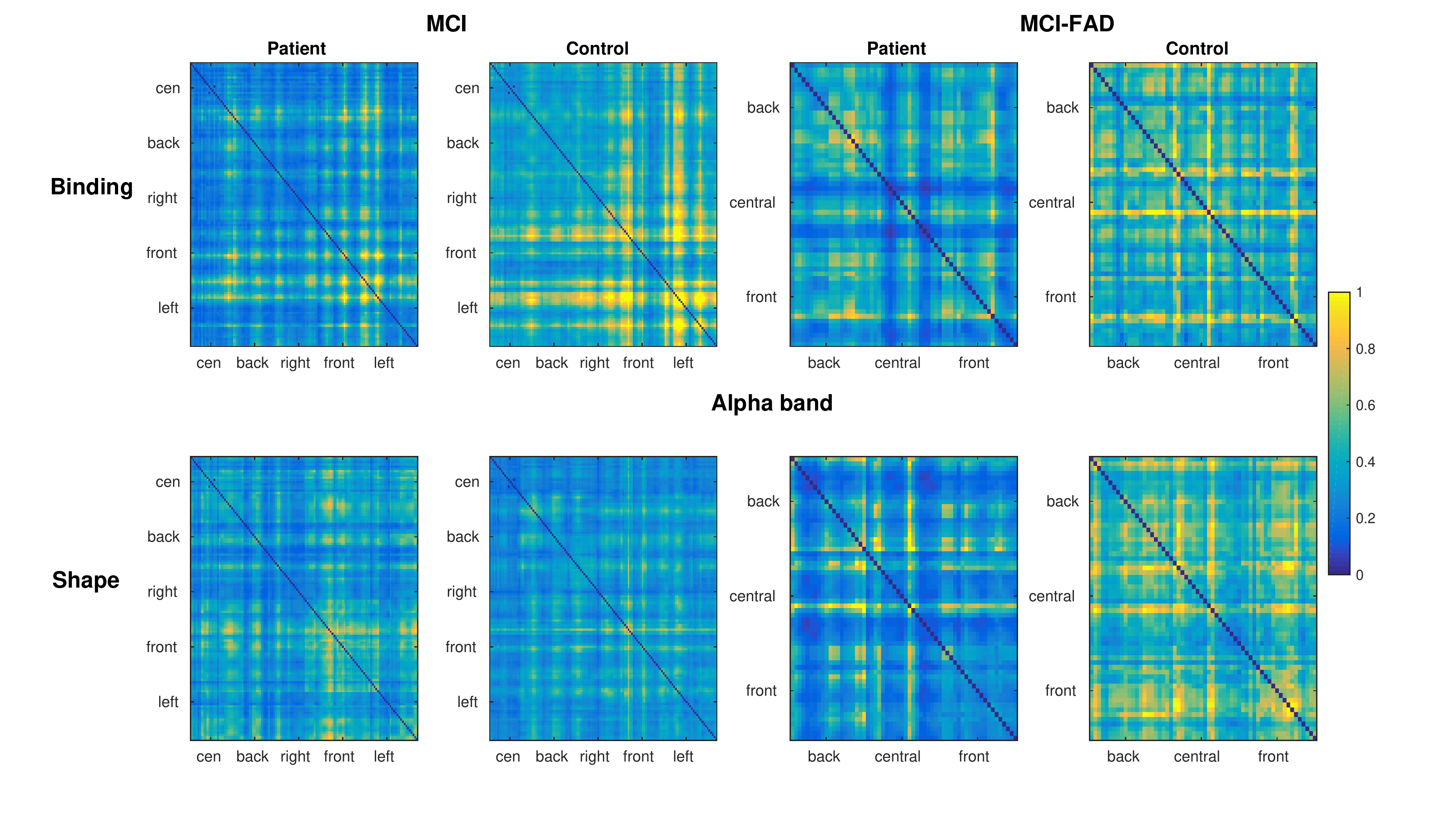}
	\caption{Connectivity map for the tensor-PLI measure and the MCI (left) and MCI-FAD (right) datasets in the alpha band for $R=8$. Each point indicates the coupling between two electrodes which have been grouped into five regions over the scalp.}\label{fig:AR8}
\end{figure*}

\begin{figure*}[htbp]
	\centering    
	\includegraphics[width=18cm]{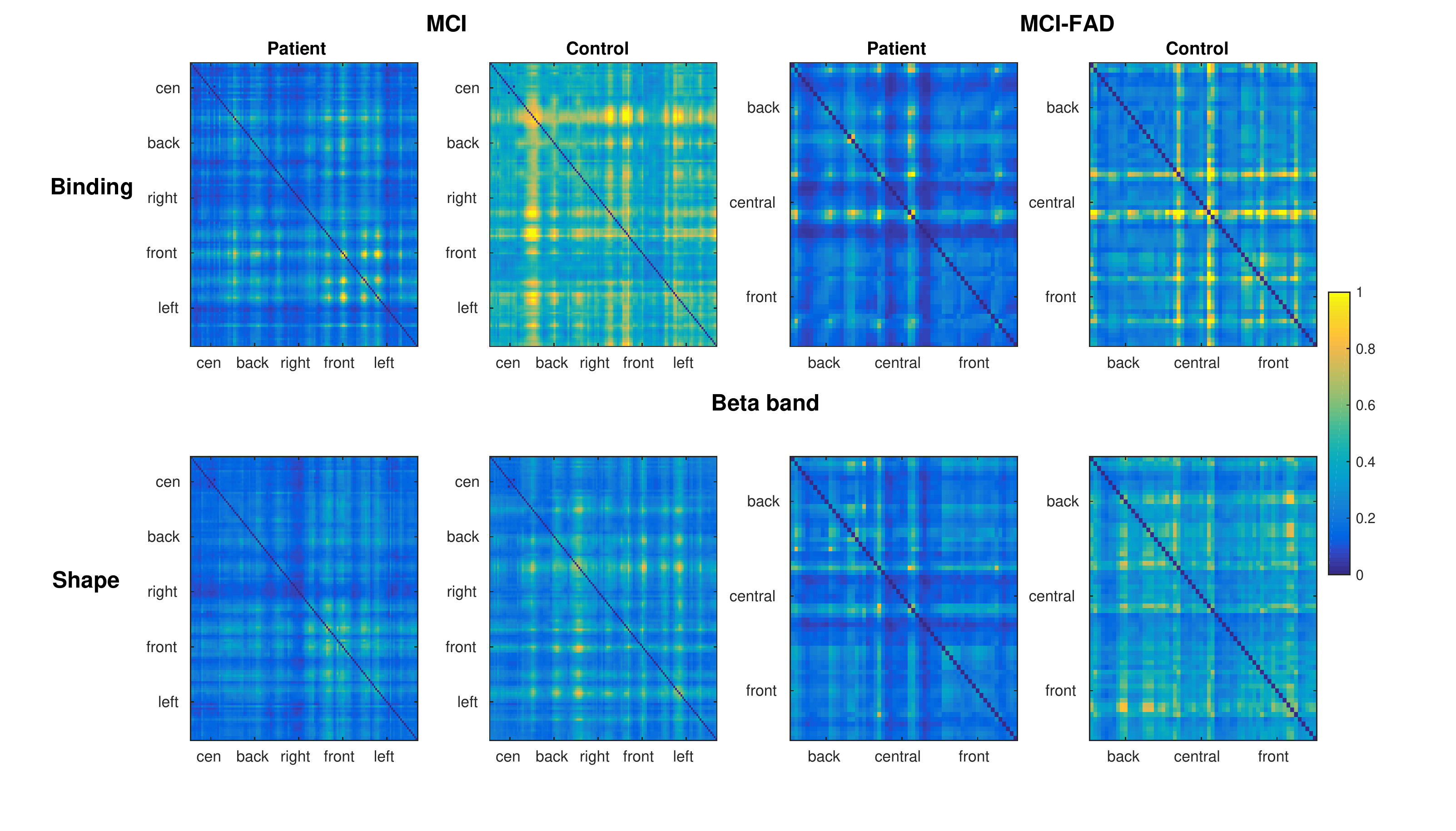}
	\caption{Connectivity map for the tensor-PLI measure and the MCI (left) and MCI-FAD (right) datasets in the beta band for $R=8$. Each point indicates the coupling between two electrodes which have been grouped into five regions over the scalp.}\label{fig:BR8}
\end{figure*}

\begin{figure*}[htbp]
	\centering    
	\includegraphics[width=18cm]{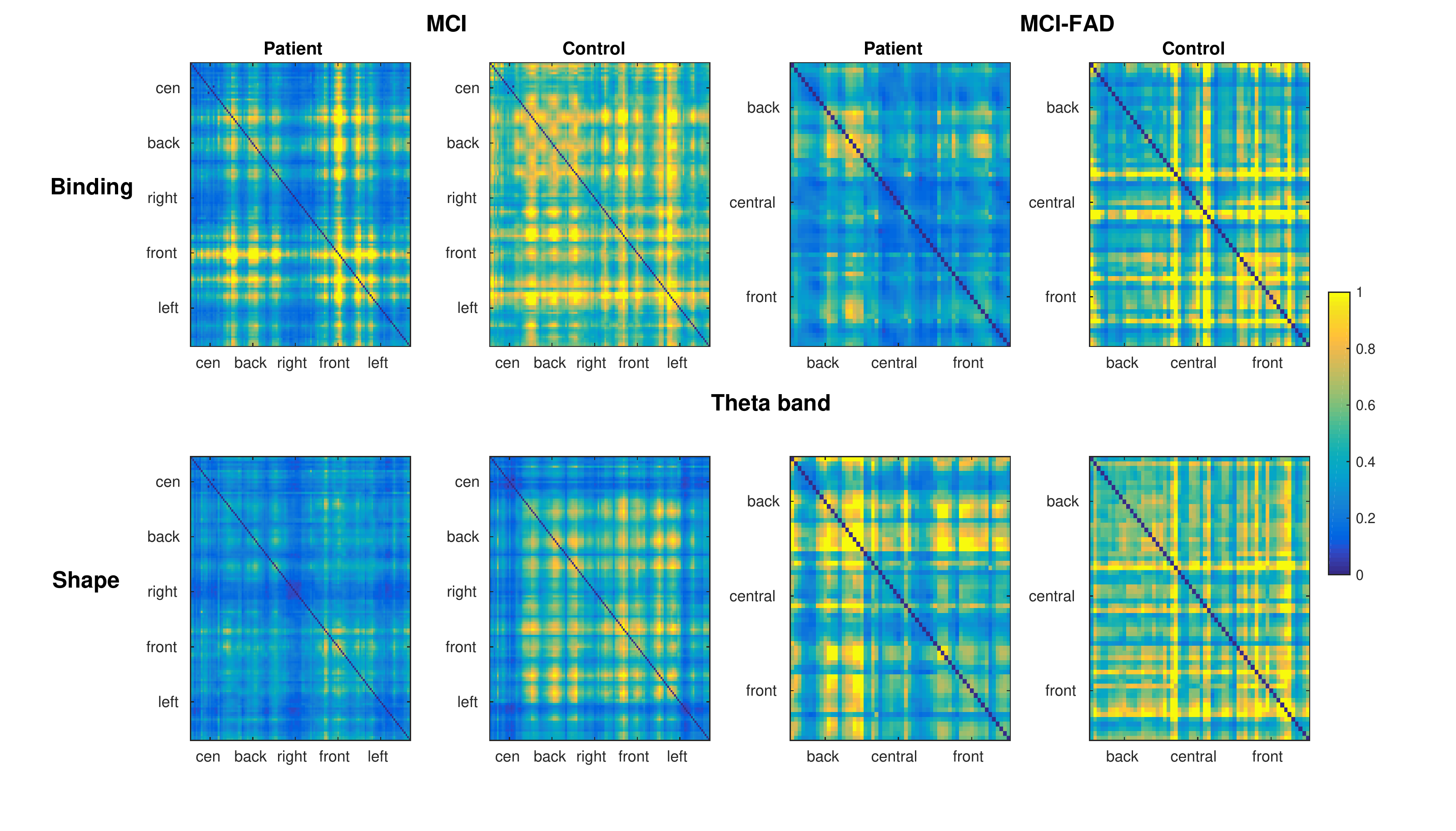}
	\caption{Connectivity map for the tensor-PLI measure and the MCI (left) and MCI-FAD (right) datasets in the theta band for $R=8$. Each point indicates the coupling between two electrodes which have been grouped into five regions over the scalp.}\label{fig:DR8}
\end{figure*}

\begin{figure*}[htbp]
	\centering    
	\includegraphics[width=18cm]{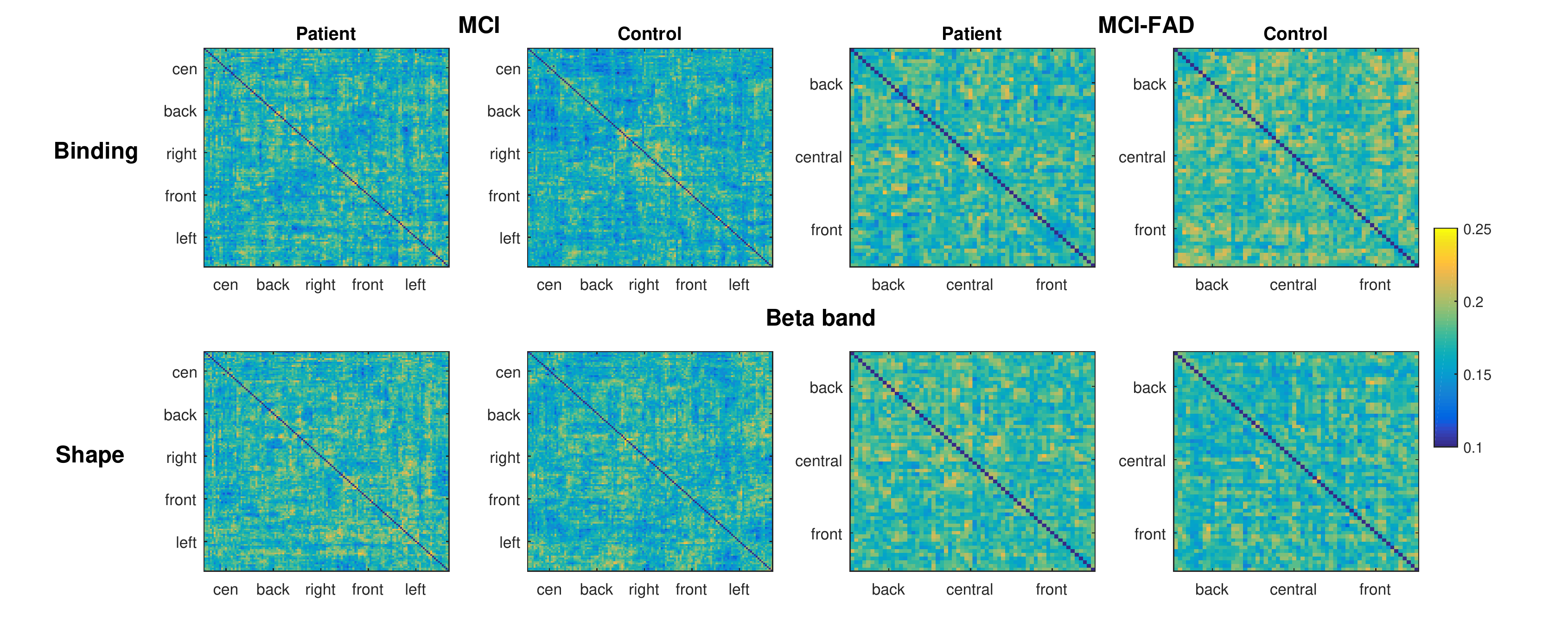}
	\caption{Connectivity map for the  standard PLI measure and the MCI (left) and MCI-FAD (right) datasets in the beta band. Each point indicates the coupling between two electrodes which have been grouped into five regions over the scalp.}\label{fig:pli}
\end{figure*}

\begin{figure}[htbp]
	\centering    
	\includegraphics[width=8cm]{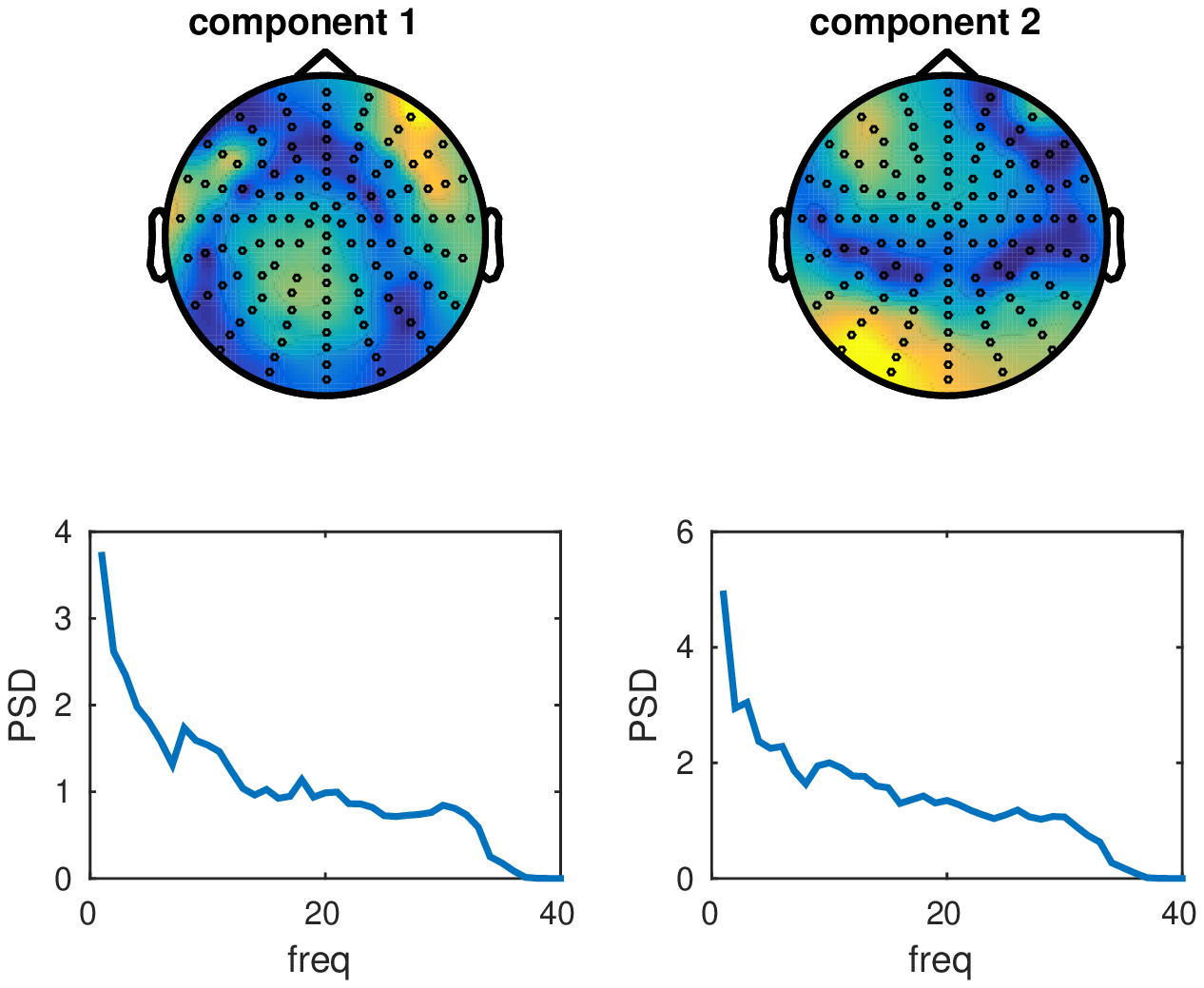}
	\caption{Example source pair obtained by PARAFAC2. Top row shows the spatial topography of the two sources while the bottom row shows their frequency profiles. These two components had a connectivity strength of 0.25. }\label{fig:ex}
\end{figure}

For both datasets, we applied a $R=8$ component complex PARAFAC2 algorithm and calculated the connectivity map on the pair of sources with the highest connection value on the alpha (8-12Hz), beta (13-20Hz) and theta (4-7Hz) frequency bands, see equation \ref{eq:conn}. In Figures \ref{fig:AR8}, \ref{fig:BR8} and \ref{fig:DR8} we show the connectivity map obtained with tensor factorisation on the alpha and beta bands. Figure \ref{fig:pli} shows the connectivity map obtained through the standard PLI measure for beta bands only. An example source pair is shown in Figure \ref{fig:ex}.

We also calculated the power ratio (PR) similar to the benchmark dataset. The differences in PR values between groups, task and frequency band were not significantly different with a mean of $0.35$ for the MCI and $0.32$ for the FAD group. 

There were statistically significant differences with an unpaired t-test in the coupling metric, Equation \ref{eq:tpli}, between the MCI group and the matched controls ($p<0.05$) for both the binding and the shape tasks with the MCI group exhibiting greater coupling. No differences were found for the MCI-FAD group and the matched controls.
In terms of the power weighted metric of Equation \ref{eq:conn} we performed a 3-way ANOVA analysis separately for the MCI and MCI-FAD datasets. The factors were condition (patient-control), task (binding-shape) and frequency band (alpha-beta-theta), with the dependent variable being the average strength over the whole scalp. Significant differences were found in condition and frequency band for both datasets (MCI/condition: $F=8.49, p<0.01$,  MCI/band: $F=3.77,p<0.05$, MCI-FAD/condition: $F=5.24,p<0.05$, MCI-FAD/band: $F=5.11,p<0.01$) with significant differences in task for only the MCI dataset only (MCI/task: $F=6.24,p<0.05$). Interactions between task and condition revealed significant differences in the MCI group between the controls and patients for both tasks. We also tested the correlation between the coupling metric and power weighted metric with significant anti-correlation ($r=-0.5, p<0.01$) for the shape task and the patient group for both datasets.
%

\section{Discussion}\label{sec:discussion}

Complex tensor factorisation enables the estimation of brain connectivity in scalp source space and its efficacy was shown in both a benchmark EEG dataset and two EEG datasets comprising of Alzheimer patient and control data. Performing tensor factorisation in the complex domain enables the calculation of brain connectivity metrics since we obtain estimates of the sources phase information. The theoretical justification of using PARAFAC2 as the tensor factorisation model was demonstrated in sections \ref{sec:EEG} and \ref{sec:tensor}. In the former we describe the way neural processes are described in the complex domain and show their dependency on frequency and trial for both phase- and non phase-locked sources. 

Initially we measured the performance of the algorithm in the benchmark dataset, where we show in Figure \ref{fig:sim}, that the complex PARAFAC2 algorithm is able to decompose the underlying data into components that reflect the true activity better than complex PARAFAC while exhibiting good detection performance. Using such a benchmark dataset we were able to generate EEG signals of varying power ratios and evaluate the performance for a wide range. We measure three different performance metrics describing various aspects of the algorithm. In Figure \ref{fig:sim} we showed the perfomance in terms of explained variance for various PR values and number of components $R$. This value describes how well the PARAFAC2 model can fit the data. Moreover, the efficacy in terms of extracting the desired sources is estimated in two ways. Firstly, by computing the CONN metric. This way we obtain an objective measurement of how powerful the signal has to be to be able to reliably extract its phase information. As observed, increasing the PR increases the CONN for PARAFAC2 but not for PARAFAC. Secondly, we estimate whether the location of the highest connected sources correspond to their true location. For all cases the increased efficacy with increased PR and $R$ was evident. A ceiling level was reached at around 0.6 PR where $R=2$ and $R=8$ runs of PARAFAC2 performed equivalently. Further increase in $R$ improves the results but the uniqueness requirements are violated as obtained through empirical results for PARAFAC2 \cite{Kiers1999,Bro1999}. The obtained values for CONN and LOC can be contrasted to those obtained in \cite{Haufe} where CONN=0.51 and LOC=0.5 over 100 datasets for PR in the range of [0.2-0.9]. However, their work assumes prior knowledge of the oscillatory sources' frequency bands; such information is not required in our algorithm. Furthermore, based on that prior knowledge of the frequency band a decision for the CONN or LOC was made in \cite{Haufebb} only when the SNR is $50\%$ higher than the background activity. Essentially, this renders the simulations for $PR\geq 0.7$. In this work, we aim to describe the performance of PARAFAC and PARAFAC2 without any heuristic measures and establish a kind of benchmark that future studies can be compared to.

In the real EEG dataset we demonstrate the performance of the algorithm by applying to two task based Alzheimer's datasets. The motivation behind the tests is to test if prior physiological knowledge about the Alzheimer's patients can be also obtained through complex PARAFAC2. Based on a memory task \cite{Pietto2016} the expectation is that the reduced performance of the patient groups should be explained by the loss of neural mechanisms. As seen in Figures \ref{fig:AR8}, \ref{fig:BR8} and \ref{fig:DR8}, the power adjusted connectivity metric, as defined in Equation \ref{eq:conn}, of the control groups is higher for both tasks and for the alpha, beta and theta frequency bands. In contrast, the synchronisation metric of Equation \ref{eq:tpli} was significantly higher for the MCI patient group compared to the matched controls. It has been demonstrated that MCI patients exhibit higher synchronisation in MEG studies \cite{Buldu2011} and functional MRI studies \cite{Dickerson2005}. The disparity between lower power but higher synchronisation between the patients and controls of the MCI group but not the MCI-FAD group can be explained firstly by noting that the average age of the MCI group is much higher (73) than the MCI-FAD group (44). The Scaffolding Theory of Cognitive Aging and Decline (STCA) \cite{Goh2009,Cabeza2002} describes the compensatory mechanisms that are recruited by the brain to alleviate age related cognitive decline. In memory tasks, MCI patients exhibit increased connectivity between brain areas as aging progresses and compared to controls \cite{Cabeza2002}. This is also shown in this study where increased connectivity is found in the MCI patient group only as compared to the age matched controls due to the larger cognitive decline corroborating the STCA theory. Lastly, the binding task consisted of more distributed brain activations for the binding than the shape task which provides further evidence that the binding tasks requires brain connectivity between different brain areas since the shape and colour binding process invokes separate brain centres \cite{Parra2014}. 

\section{Conclusions} \label{sec:conclusions}

In this work we showed that complex tensor factorisation based on complex PARAFAC2 is suitable for EEG connectivity estimation and superior to the complex PARAFAC. By establishing that EEG data follow the PARAFAC2 model in the complex domain a tensor factorisation algorithm was successful in a benchmark and two Alzheimer's EEG datasets.

This work has the following implications. Firstly, the use of high order methods such as tensor factorisation is suitable for extracting coupled sources. And secondly, performing tensor factorisation in the complex domain allows for the connectivity information present in the data to be optimally exploited. Importantly, the decoupling of spatial, spectral, and synchronisation behaviour of networks is not possible with traditional connectivity metrics and this work aims to bridge that gap. This works opens up a new possibility in the inspection of EEG connectivity.

Future work entails modification of the PARAFAC2 constraint to better facilitate the EEG data generation model. In that light, different data models will be considered such as phase/non-phase locked, event related potentials, and the presence of multiple coupled source pairs. We also to plan to validate the algorithm on a longitudinal AD dataset of more than 100 patients over 5 years.

\section{Acknowledgments}

We wish to thank Agust\'{i}n Ib\'{a}\~{n}ez for the provision of the dataset \cite{Pietto2016}.

\bibliographystyle{IEEEbib}
\bibliography{../citations}
\end{document}